\documentclass[journal]{IEEEtran}

\usepackage{cite}
\usepackage{amsmath,amssymb,amsfonts}
\usepackage{algorithmic}
\usepackage{graphicx}
\usepackage{textcomp}
\usepackage{xcolor}
\usepackage{booktabs} 
\usepackage{gensymb}
\usepackage{array}
\usepackage{chemist}
\usepackage{tcolorbox}
\usepackage{hyperref}
\usepackage[margin=0.8in]{geometry}
\usepackage{lscape}
\usepackage{adjustbox}
\usepackage{indentfirst}
\usepackage{multicol, blindtext}
\usepackage{pifont}
\usepackage{lettrine}
\usepackage{pdfpages}
\usepackage{multirow}
\usepackage{tabularx}
\usepackage[compatibility=false]{caption}
\usepackage{makecell}
\usepackage{float}
\usepackage{caption}
\usepackage{enumitem}
\newcolumntype{Y}{>{\raggedright\arraybackslash}X}

\PassOptionsToPackage{hyphens}{url}
\usepackage{hyperref}

\usepackage{soul}

\usepackage{ragged2e}

\hypersetup{%
colorlinks=true,
linkcolor=black,
citecolor=black,
urlcolor=black}

\usepackage{verbatim}

\newcommand{\cmark}{\ding{51}}

\definecolor{forestgreen}{rgb}{0.13, 0.55, 0.13}

\begin{document}

\title{ Counting own goals: High-level assessment of the economic relationship between the ICT and the Oil and Gas sectors and its environmental implications}

\author{
\IEEEauthorblockN{Gauthier Roussilhe\textsuperscript{1,*}, Béatrice Dromard\textsuperscript{1}, Srinjoy Mitra\textsuperscript{2}}

\IEEEauthorblockA{\textsuperscript{1}\textit{Hubblo}, Paris, France}
\IEEEauthorblockA{\textsuperscript{2}\textit{School of Engineering}, Edinburgh}
}

\maketitle

\renewcommand{\thefootnote}{\fnsymbol{footnote}}
\footnotetext[1]{Corresponding author: gauthier.roussilhe@hubblo.org}
\renewcommand{\thefootnote}{\arabic{footnote}}

\maketitle

\begin{abstract}

The ICT sector has been one of the most successful and fastest-growing industry in history. While the environmental issue in this sector has mainly been addressed by assessing its footprint and, to a lesser extent, its avoided emissions or net impacts, the additional emissions from the digitalization of carbon-intensive activities, such as the Oil and Gas (O\&G) sector, have rarely been discussed. By doing so, we have forgotten to count the own goals conceded over more than 20 years in the troubled relationship between the ICT and the O\&G sector. Using input-output analysis and economic data ranging from 2000 to 2022, we observe that on average 2\% of the annual financial flows from the ICT sector are directed towards the Oil and Gas sector. Considering the significant growth of the ICT sector during this time, O\&G companies now spends a massive amount on ICT products in absolute terms. It also appears that in 2022, for each dollar going from the ICT sector to the renewable and nuclear energy industry, more than \$4 go to the O\&G industry. In addition, we also provide a classification of digital activities in the O\&G sector to facilitate environmental assessments and present two case studies estimating potential added emissions from the digitalization of oil activities. Finally, looking at the immense growth in generative AI, we provide an exploration of causal links between the current success of GPU technology and its intricate early relationship with the O\&G sector. This article lays the groundwork for defining the nature of the relationship between ICT and O\&G, which predates the current hype surrounding generative AI. We  provide the analytical elements needed to begin estimating the added emissions from the digitalisation of O\&G.

\end{abstract}

\textbf{Data Availability Statement:} The data that supports the findings of this study are available in the supporting information of this article.

\textbf{Conflict of Interest:}
The authors have no conflict of interest to disclose. 

\section{Introduction}
For more than fifty years, the positive environmental effects of the ICT (or digital) sector have been studied by public institutions\cite{information1996networks}, tech companies\cite{GSMA2019EnablementEffect}, civil society\cite{pamlin2002sustainability} and the research community\cite{hilty2006relevance, coroamua2019digital}. We are getting a better understanding of the environmental impact of this sector \cite{furberg2026guidelines} and the potential contribution of digitalization towards ecological sustainability\cite{bieser2023review}. While digitalization has spread to all areas of the modern economy and most economic activities, the Oil and Gas (O\&G) industry is also heavily dependent on it. However, despite the present-day efforts of most industrial sectors to mitigate their environmental impact, the O\&G industry has been continuously extracting an increasing amount of hydrocarbons, at a time when a substantial fossil fuel phase-out is imperative.  At the intersection of these two phenomena, there is almost no literature on the contribution of ICT to the growth of the O\&G industry. The extensive scholarship on the digitalization of oil and gas activities rarely translates efficiency gains and increased production into environmental impacts\cite{saputelli2024success}.

To understand the overall environmental impact of O\&G digitalization, we must consider potential avoided and added emissions. While the O\&G sector has nearly doubled its production capability since 1990 (from $\sim$55,000 TWh to $\sim$100,000 TWh)\cite{EI_StatisticalReview_2024}, yet, the popular consensus on ICT's role in this growth seem to tip in favor of avoided emissions. For example, the latest IEA report on AI and energy has in depth calculations on the potential for AI to achieve CO\textsubscript{2} reductions, but devoted only a brief section to AI-induced additional productivity in the Oil and Gas sector. This analysis entirely ignored such possibilities of added emissions when calculating the potential for AI-induced emission reduction\cite{iea2025energyai}. Meanwhile, the CEO of Saudi Aramco, the world’s largest oil company, recently estimated that AI could double oil-well productivity\cite{arabweekly2025aramcoai}. To reset the scales, we must establish a foundation for understanding the relationship between digitalization, the Oil and Gas sector, and their intertwined environmental impact.

This paper proposes to establish a conceptual and historical basis for understanding the co-evolution of the ICT sector and the O\&G sector. This reconstruction is structured on three levels:
\begin{itemize}
  \item At a macroeconomic level, how much ICT has actually been consumed by the O\&G sector since the early days of its digitalization?
  \item At a market level, which ICT products and services have been and are being used by the O\&G industry, and for what intended effects?
  \item At the environmental level, how does the use of digital technologies by the O\&G industry be translated in environmental impacts, even with minimal data access?
\end{itemize}

\section{Methodology}
We used parallel methodologies to address the three research questions raised in this article: Input-Output Analysis and a focused literature review of the digitalization of Oil and Gas activities.

\subsection{Input-Ouput Analysis (IOA)}
First, we used the Input-Output model \cite{suh2009handbook} to define the macroeconomic relationship between the ICT sector and the O\&G sector. Input-Output tables document the inputs from an economic sector used by another sector in the production of its own outputs. Input-Output Analysis has been used in  various contexts: economic planning by institutions, economic analysis in research, or even environmental analysis with the availability of environmental data in certain IO databases.

A regional dimension can be added to the Input-Output tables, documenting the geographical origin of inputs used in another region by another economic sector. The input-output tables used in our research come from the EXIOBASE3 database\cite{stadler2018, EXIOBASE3}, which is constituted of such \textit{Multi-Regional Input-Output tables} (MRIO). We  only parsed the transactions tables, documenting the economic relations between regions and sectors. The tables present the absolute values (in millions of euros) of the inputs received by a sector from another sector, primarily materials and services utilized within the value chain of an economic sector. The monetary value employed in Input-Output transaction tables for the quantification of these inputs is necessary for the standardization of data. In the event of the actual materials being utilized in the transaction tables (for instance, one computer from China being employed in the production of one barrel of oil in Saudi Arabia), the parsing and comprehension of such data would prove to be a rather arduous task. The database maintained do not encompass financial transfers between sectors, such as financial investments flowing from one company to another within the same sector.


Based on recent Input-Output Analysis work on the ICT sector\cite{charpentier2023estimating, charpentier2025comparing}, we have defined the categories of economic activities representing the ICT sector and those representing the O\&G sector.

It is important to note that EXIOBASE3 uses the categories defined in the International Standard Industrial Classification\cite{un_isic_rev4_2008} in an aggregated manner. Consequently, we cannot separate certain flows that do not form part of the ICT sector, such as printing or office machinery. What we refer to here as the ICT sector for the sake of simplicity correspond to broader scopes than those traditionally defined in studies on the environmental footprint of these sectors, i.e. the ICT sector, some Entertainment and Media activities, and few out-of-scope activities. The category \textit{Manufacture of office machinery and computers} aggregates not only the manufacture of computers and peripherals, but also that of cash registers and photocopiers, among others. The category \textit{Manufacture of radio, television and communication equipment and apparatus} primarily aggregates the manufacture of radios and televisions. The category \textit{Post and telecommunications} covers telecommunications activities (wired, wireless, satellite) and television and radio broadcasting. The category \textit{Computer and related activities} covers the sale of computers, peripherals, telecommunications equipment and software. Finally, the category \textit{Publishing, printing and reproduction of recorded media} covers printing activities as well as the reproduction of recorded media such as high-resolution image files, etc.

Regarding the O\&G sector, we excluded coal-related activities due to the different nature of their industrial practices with oil and natural gas extraction and refining. In order to provide a basis for comparison, we have also included inputs from ICT going to the renewable and nuclear (R\&N) energy sectors. This classification is summed up in Table \ref{tab:sectors}.

\begin{table}[t]
\centering
\footnotesize
\setlength{\tabcolsep}{6pt}
\renewcommand{\arraystretch}{1.2}

\begin{tabularx}{\columnwidth}{lX}
\toprule
\textbf{Sector} & \textbf{Activities} \\
\midrule

\textbf{ICT sector} &
\begin{itemize}[leftmargin=*, noitemsep, topsep=0pt]
\item Manufacture of office machinery and computers
\item Manufacture of radio, television and communication equipment and apparatus
\item Post and telecommunications
\item Computer and related activities
\item Publishing, printing and reproduction of recorded media
\end{itemize} \\

\textbf{O\&G sector} &
\begin{itemize}[leftmargin=*, noitemsep, topsep=0pt]
\item Extraction of crude petroleum and related services (excluding surveying)
\item Extraction of natural gas and related services (excluding surveying)
\item Extraction, liquefaction, and regasification of petroleum and gaseous materials
\item Petroleum refining
\item Electricity generation from gas
\item Electricity generation from petroleum and derivatives
\item Manufacture and distribution of gaseous fuels
\end{itemize} \\

\textbf{R\&N sector} &
\begin{itemize}[leftmargin=*, noitemsep, topsep=0pt]
\item Mining of uranium and thorium ores
\item Processing of nuclear fuel
\item Electricity generation: nuclear, hydro, wind, biomass, and solar photovoltaic
\end{itemize} \\

\bottomrule
\end{tabularx}

\caption{Sectoral classification of activities based on ISIC Rev.4}
\label{tab:sectors}
\end{table}

The economic flows between these sectors were mapped over a period from 2000 to 2022. The initial year was chosen as the starting point for observing the transformations linked to the growing development of digital technologies in the global economy. The end year corresponds to the last year available in EXIOBASE3 at the time of the study.

\subsection{Focused literature review}

Following the macroeconomic perspective given by the IOA, the aim of this focused literature review is to provide a high-level overview of the integration of digital technologies and services in the O\&G industry over the last few decades. The literature review focused primarily on the specialist journals of the Society of Petroleum Engineers (SPE): the Journal of Petroleum Technology (JPT), OnePetro and the SPE Journal. We selected review papers that provided both a historical overview of the sector’s digitalisation and of the specific technologies employed. Two review articles offer such a historical perspective: Cramer\cite{cramer2008back}, covering the period from the 1960s to the early 2000s, and Saputelli et al.\cite{saputelli2024success}, covering the period from the 2000s to 2020.

This first review is supplemented by industry literature provided by major oil and gas companies on the digitalization of their activities\cite{bp2018technology, aramco_digital_technologies} and by major technology companies on the services they provide to the O\&G industry\cite{hedge2019securing}. Finally, recent reports from consulting firms and international organizations on the digitalization of this industry are also taken into account\cite{pwc2021supercharging, geissbauer2019digitizing}. We specifically looked for articles and reports that set out as clearly as possible the expected outcomes of implementing a digital solution, and which provided a quantification of expected outcomes.

The aim of this literature review is to identify the digital products and services purchased by the O\&G sector, their long-term trends, and the reasons motivating these purchases. Such an analysis provides a better understanding of the economic flows modelled by the IOA at a macro level. It also enables a first basis to inventory and classify the digital technologies used in the O\&G industry value chain and the desired effects of integrating these technologies.

\subsection{Examples of ICT solutions used in the O\&G sector and high-level carbon footprint assessment}

Based on this literature review, we aim to select examples of ICT solutions operating at different levels of the O\&G sector’s value chain and providing an estimate of the gains achieved (increased production, improved recovery, cost reduction, downtime reduction). This selection criterion is motivated by the aim to analyze the potential environmental consequences of digital solutions in this sector, particularly with a perspective of assessing net impacts as defined by ITU-T L.1480\cite{itu20221480}. For example, an increase in production—whether through improved efficiency or optimization—brought about by a digital solution typically results in additional barrels and thus higher emissions, constituting a direct rebound effect. Savings in time or money may allow for partial reinvestment in more or less carbon-intensive activities, constituting an indirect rebound effect that needs to be quantified using specific methods\cite{ITU_LSup54_2022}.

The selection of examples should therefore test the classification of digital solutions in the O\&G sector and their intended effects, whilst also providing illustrations of how additional emissions can be quantified.

\graphicspath{{./plots}}

\section{Results}

\subsection{Input-Output Analysis of ICT and O\&G sectors}

\begin{figure*}[h!]
\centering
\includegraphics[width=1\linewidth]{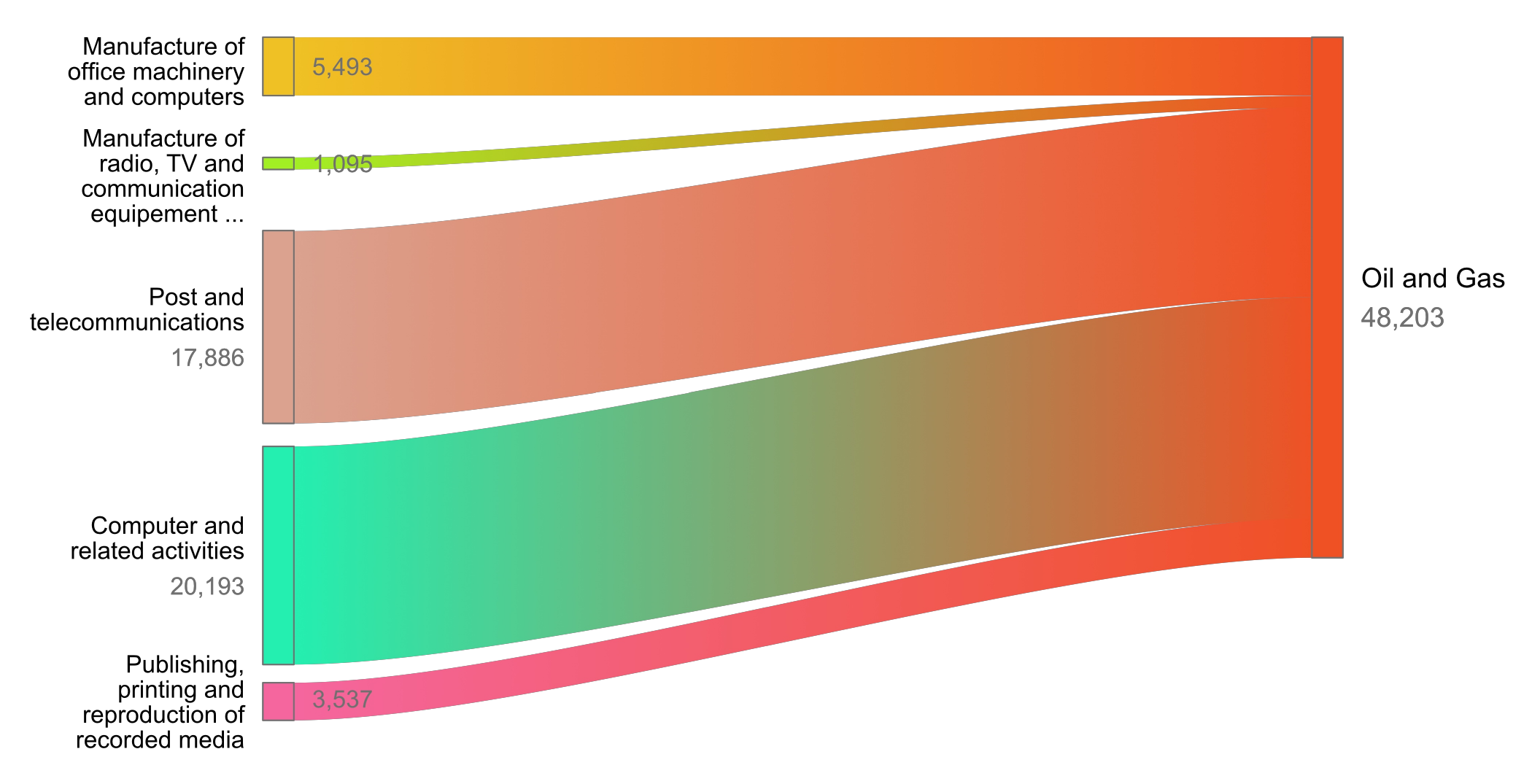}
\caption{Global ICT sector inputs to global O\&G sector, in 2022, in millions of Euros}
\label{fig:sankey_ict_to_og_2022}
\end{figure*}

Figure \ref{fig:sankey_ict_to_og_2022} shows the inputs from the ICT sector to the O\&G sector in 2022. This corresponds to the economic flows from ICT activities, as defined by EXIOBASE3, to the O\&G sector. These flows include purchases, sales, transactions or any economic interdependency between the two sectors. Two ICT activities (Post and Telecommunications, Computer and related activites) represent the main inputs flowing to the O\&G sector. As a reminder, the \textit{Post and Telecommunications division} mainly corresponds to fixed, mobile and satellite telecommunications activities. The \textit{Computer and related activities} division includes software development and support, the creation and management of computer systems and data processing facilities, and any other related activities.

\begin{figure}[h]
\includegraphics[width=8cm, height=8cm]
{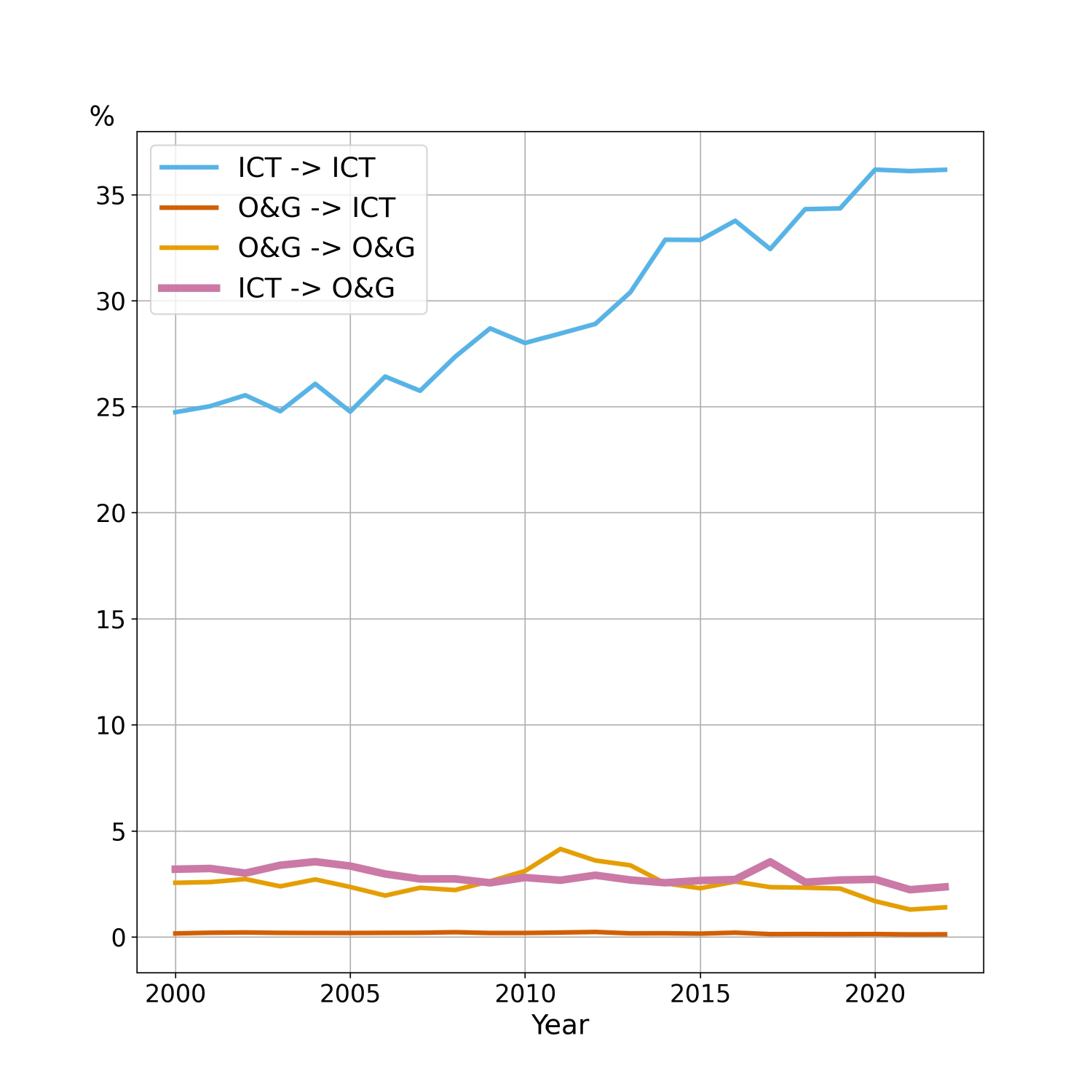}
\caption{Evolution of O\&G and ICT shares across O\&G and ICT, from 2000 to 2022, in relative value}
\label{fig:sankey_ict_og}
\end{figure}

A similar analysis of the ICT sector shows in Figure  \ref{fig:sankey_ict_og} that, on average, from 2000 to 2022, 2\% of annual ICT inputs went to the O\&G sector, even when the total inputs increased in absolute value. The ICT sector showed a large and increasing amount of endogenous inputs (transactions between ICT companies/activities) while inputs from O\&G to ICT remain relatively weak (0.4\% on average over the period).

\begin{figure}[h]
\includegraphics[width=8cm, height=8cm]{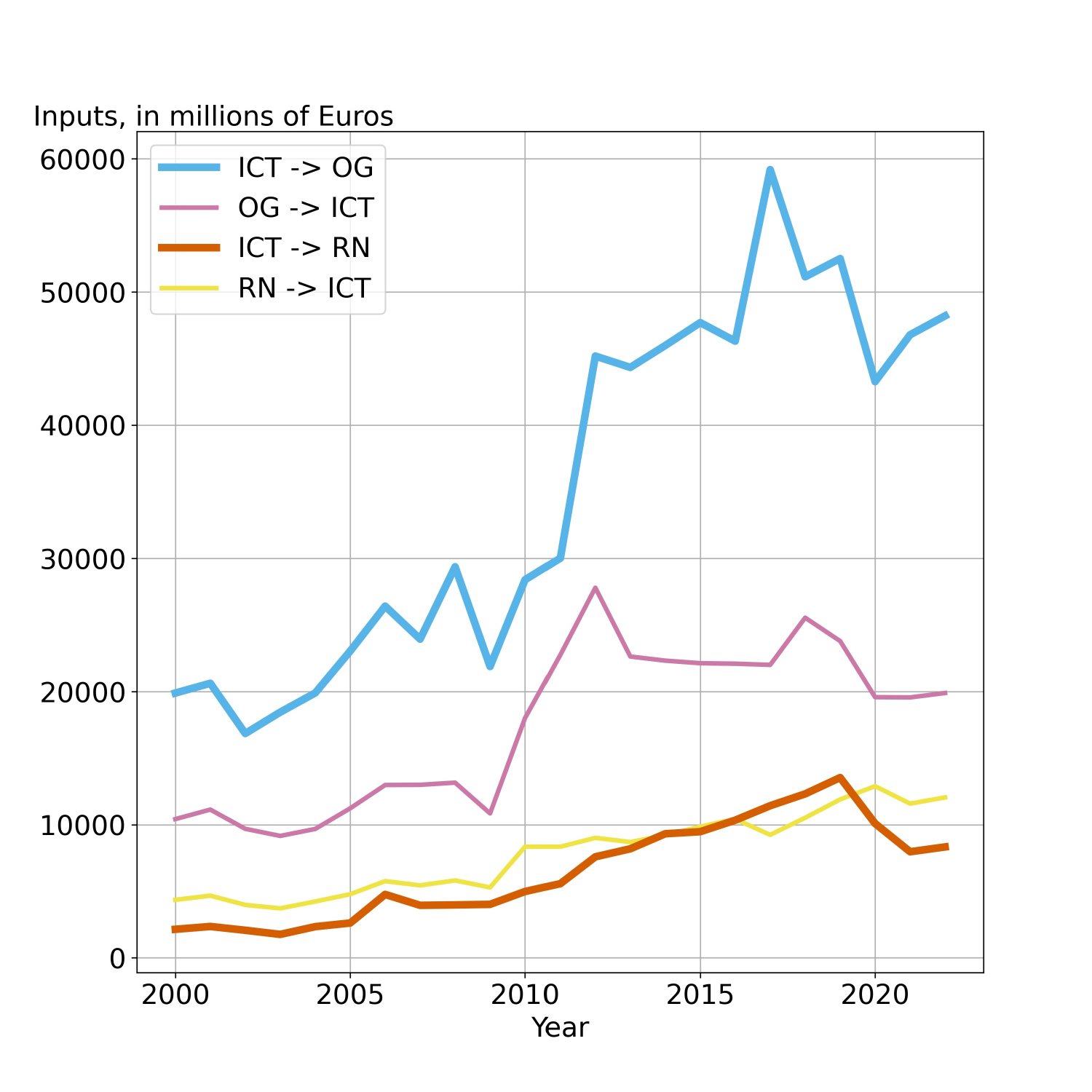}
\caption{Evolution of global ICT sector inputs to global O\&G sector, from 2000 to 2022, in absolute value}
\label{fig:inputs_totals_00_22}
\end{figure}

In Figure \ref{fig:inputs_totals_00_22}, we aggregated the ICT sector as total inputs going to the global O\&G sector, along with the total O\&G inputs going to ICT and the total ICT inputs going to the Renewable and Nuclear energy sectors (R\&N), as a reference point.

\begin{figure}[h]
\includegraphics[width=8cm, height=8cm]{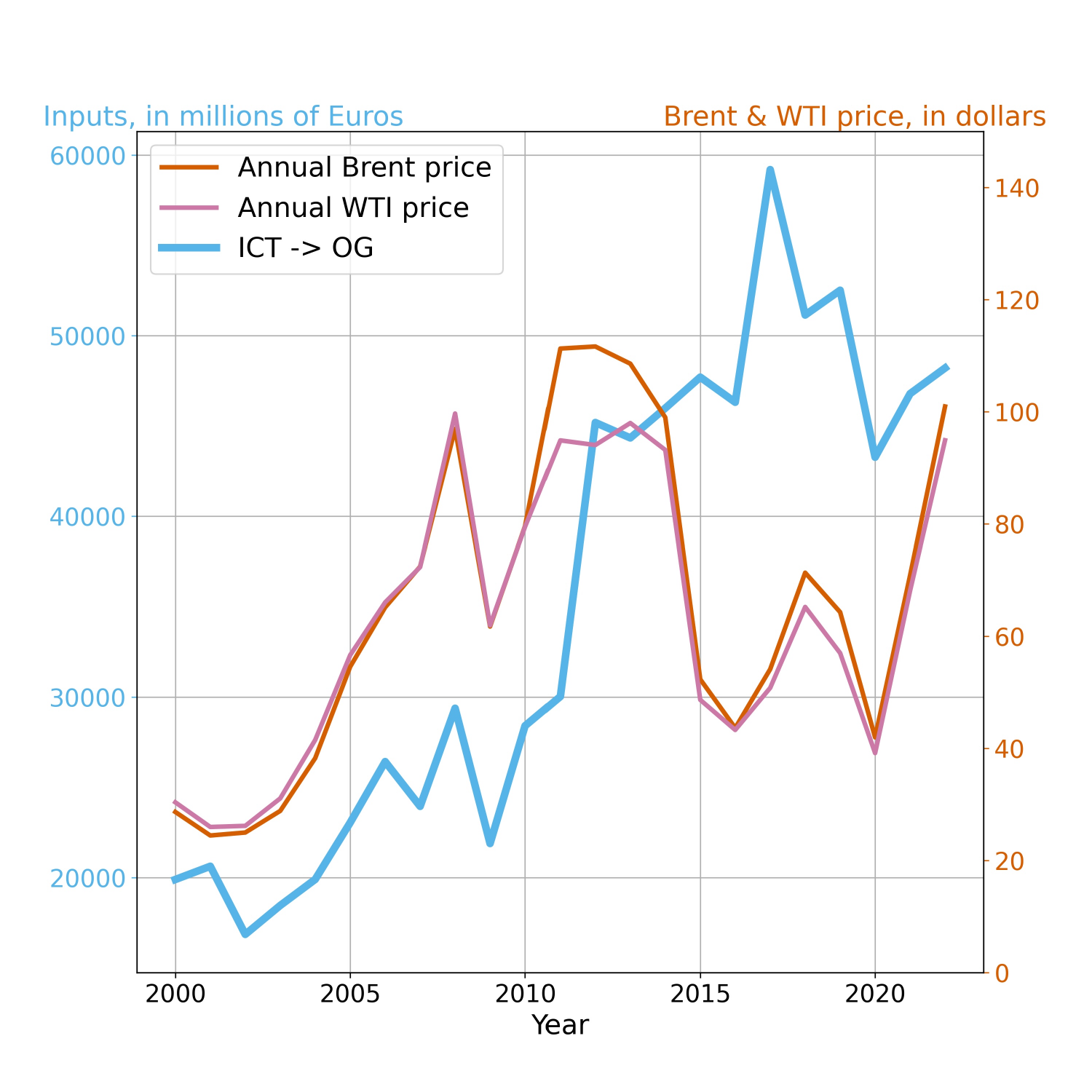}
\caption{Evolution of global ICT sector inputs to global O\&G sector, from 2000 to 2022 in absolute value, with annual oil prices}
\label{fig:inputs_total_oil_00_22}
\end{figure}

Figure \ref{fig:inputs_total_oil_00_22} shows the same ICT to O\&G line chart as figure \ref{fig:inputs_totals_00_22}, but with additional comparison with the Brent and WTI benchmarks for annual oil prices. The sharp decline in the oil prices at the beginning of the 2010’s does not seem correlated to a decline in ICT inputs to the global O\&G sector. On the contrary, there is a growth in ICT inputs to the O\&G sector around the same time. According to Al-Rbeawi\cite{al2023review}, the rapid fall in oil prices in 2014 and the supply and demand shocks spurred the sector’s digital transformation, particularly with the development of unconventional resource extraction such as shale oil, which requires hydraulic fracturing simulations and presents more complex operations. Against this backdrop, part of the sector sought to aggressively reduce operating costs and thereby improve operational efficiency, in the hope that digitalization would achieve this. Thus, the drive to reduce operating costs and the development of shale oil in the US might partly explain the highest absolute values for ICT inputs observed from 2011 to 2022.

\begin{figure*}[h]
\includegraphics[width=1\linewidth]{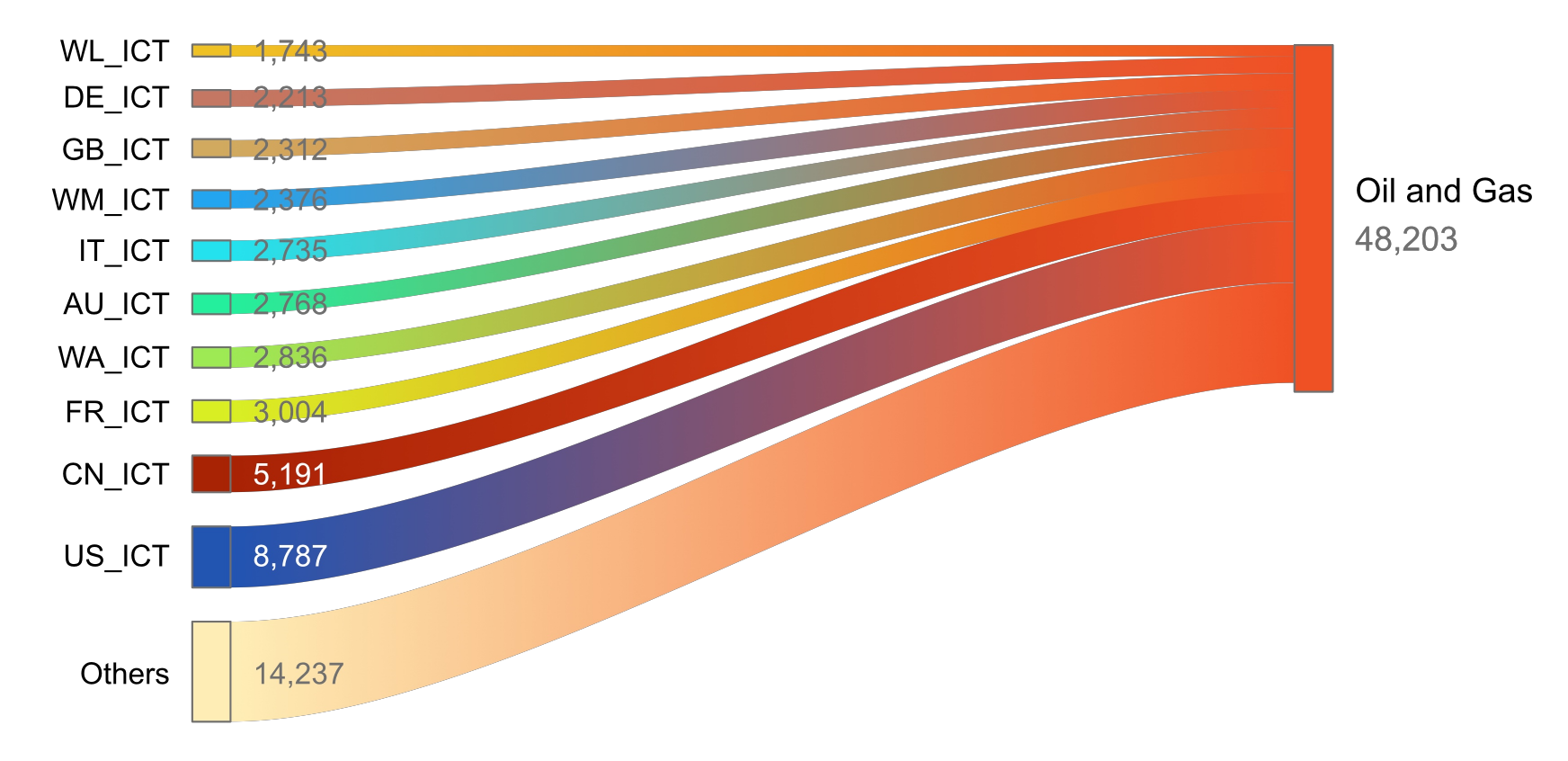}
\caption{Biggest national ICT sectors to global O\&G sector, in 2022, in millions of Euros, in absolute value}
\label{fig:main_contributors_og_22}
\end{figure*}

\begin{figure}[ht]
\includegraphics[width=8cm, height=8cm]{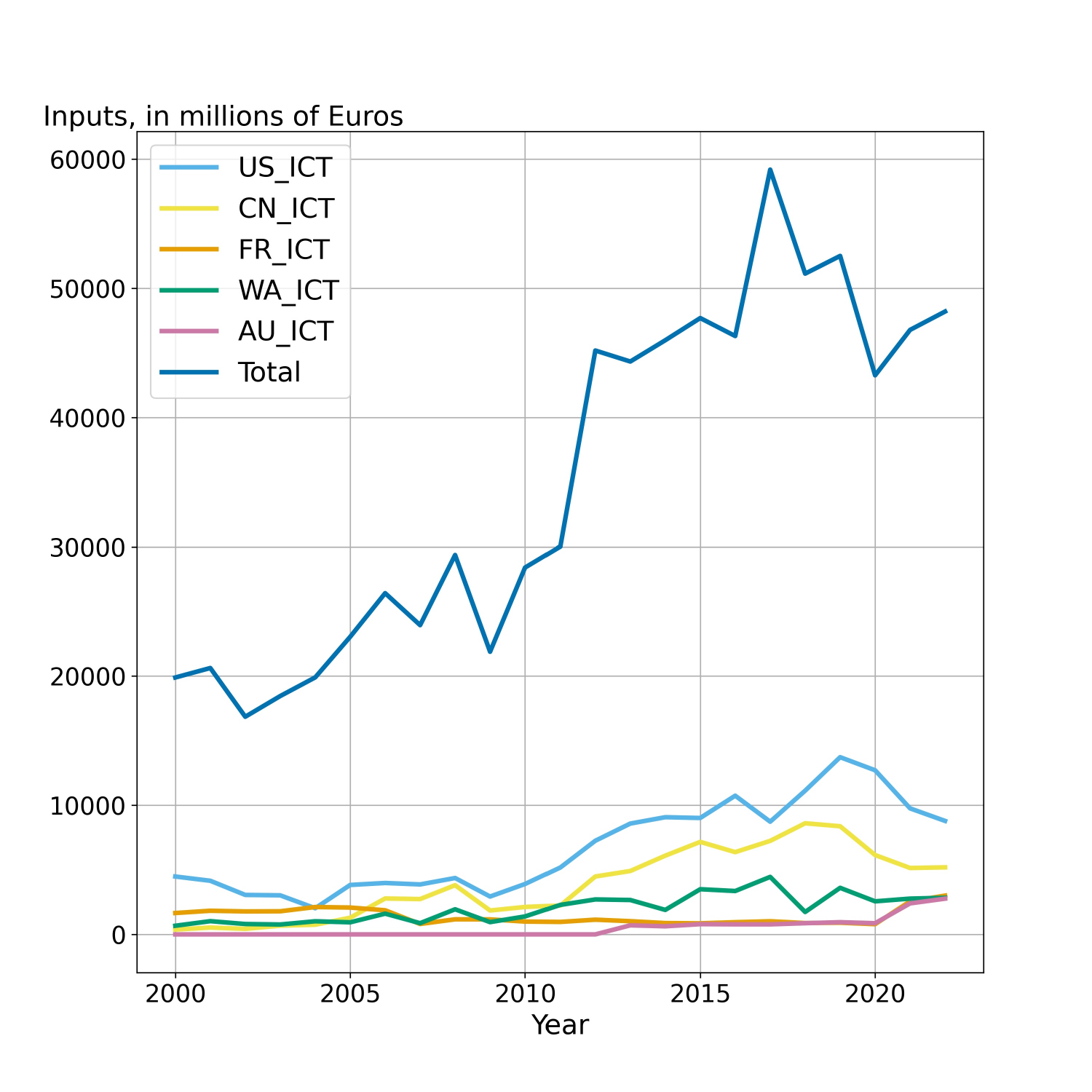}
\caption{Evolution of main ICT national contributions (2022) to global O\&G sector, from 2000 to 2022}
\label{fig:main_contributors_og_00_22}
\end{figure}

The figure \ref{fig:main_contributors_og_22} shows the main national ICT providers to the global O\&G sector in 2022. Unsurprisingly, the US and China, as the two biggest oil consumers worldwide, lead in providing ICT inputs to O\&G sector, followed by France, South-Asian countries and Australia. This provides a sanity check confirming that the biggest ICT providers in the world are also the ones most entangled with the O\&G sector. Having identified these five main contributors, this allowed us to observe the evolution of their contributions in Figure \ref{fig:main_contributors_og_00_22}.

\subsection{Digital technologies for the O\&G industry}

The literature studied makes extensive use of numerous commercial terms that are generally poorly defined or refer to different practices: automation, digital twin, artificial intelligence, Internet of Things (IoT), computer vision, big data. In order to clarify the terminology for analytical purposes, we propose a high-level classification in Table~\ref{tab:digital_oil_gas} that defines the desired functions, the hardware and digital services used, and the intended effects. This simplification aims to avoid the commercial jargon inevitably associated with this type of market and to focus solely on the essence of what one seeks to achieve when digitalizing this industry and at what level.

\begin{table*}[ht]
\centering
\setlength{\tabcolsep}{4pt}
\renewcommand{\arraystretch}{1.2}

\begin{tabularx}{\textwidth}{lll|ccccccccc|cccccc}
\toprule
\textbf{Activity} &
\textbf{Sub-activity} &
\textbf{Function} &
\multicolumn{9}{c|}{\textbf{High-level primary service / hardware}} &
\multicolumn{6}{c}{\textbf{Intended primary effect}} \\

& & &
\rotatebox{90}{HPC} &
\rotatebox{90}{DL} &
\rotatebox{90}{Sensors} &
\rotatebox{90}{Actuators} &
\rotatebox{90}{Networks} &
\rotatebox{90}{Cloud} &
\rotatebox{90}{Robotics} &
\rotatebox{90}{UAV} &
\rotatebox{90}{Mobile} &
\rotatebox{90}{Inc. Prod} &
\rotatebox{90}{Imp. Rec} &
\rotatebox{90}{Cost} &
\rotatebox{90}{Downtime} &
\rotatebox{90}{HSE} &
\rotatebox{90}{Decision} \\
\midrule

\multirow{5}{*}{Upstream}
& \multirow{2}{*}{Exploration}
& Seismic imaging and interpretation
& \cmark & \cmark & & & & & & & & \cmark & \cmark & & & & \\

&
& Reservoir simulation
& \cmark & \cmark & & & & & & & & \cmark & \cmark & \cmark & & & \\

& \multirow{3}{*}{Production}
& Drilling optimisation
& & & \cmark & \cmark & \cmark & \cmark & \cmark & & & \cmark & \cmark & \cmark & & \cmark & \\

&
& Production forecasting
& \cmark & \cmark & & & & & & & & \cmark & \cmark & & & & \\

&
& Predictive maintenance
& & & \cmark & & \cmark & \cmark & & \cmark & \cmark & & & \cmark & \cmark & \cmark & \\

\midrule

\multirow{3}{*}{Midstream}
& \multirow{3}{*}{Transportation}
& Transport optimisation
& & & \cmark & & \cmark & \cmark & & & & & & \cmark & & & \\

&
& Pipeline surveillance
& & & & & & & & \cmark & & & & \cmark & & \cmark & \\

&
& Predictive maintenance
& & & \cmark & & \cmark & \cmark & & \cmark & \cmark & & & \cmark & \cmark & \cmark & \\

\midrule

\multirow{3}{*}{Downstream}
& \multirow{3}{*}{Refining}
& Production optimisation
& & & \cmark & & \cmark & \cmark & & \cmark & \cmark & \cmark & & \cmark & \cmark & \cmark & \\

&
& Supply chain monitoring
& & & \cmark & & \cmark & \cmark & & & & \cmark & & \cmark & \cmark & \cmark & \\

&
& Predictive maintenance
& & & \cmark & & \cmark & \cmark & & \cmark & \cmark & & & \cmark & \cmark & \cmark & \\

\midrule

\multirow{4}{*}{Operations}
& \multirow{4}{*}{}
& Data management
& & & & & & \cmark & & & & & & \cmark & & & \cmark \\

&
& Business Process Management (BPM)
& & & & & & \cmark & & & & & & \cmark & & & \cmark \\

&
& Social collaboration
& & & & & & \cmark & & & & & & \cmark & & & \cmark \\

&
& Remote Operations Centers (ROC)
& & & & & & \cmark & & & & & & \cmark & & & \cmark \\

\bottomrule
\end{tabularx}

\caption{ICT activities with categorical service/hardware and intended effects. HPC: High-Performance Computing; DL: Deep Learning; UAV: Drones; Mobile: Mobile devices; Inc. Prod: Increased Production; Imp. Rec: Improved Recovery; Cost: Cost reduction; Downtime: Downtine reduction; HSE: Improved Health, Safety, Environment; Decision: Improvied Decision-Making.}
\label{tab:digital_oil_gas}
\end{table*}

Such simplification is also necessary to dispel the fog of assertions present in the literature. The effects of digitalization in the Oil and Gas industry are widely discussed, but the specific mechanisms are rarely documented. For example, a 2019 PwC report estimates that digitalizing downstream activities could reduce operating costs by 12 to 20\%, throughput by 6 to 12\%, unplanned shutdowns by 15 to 25\%, and plant efficiency by 8 to 12\%\cite{geissbauer2019digitizing}. The report does not specify which operating costs are reduced and how, nor does it discuss the difference between increased throughput and plant efficiency. Gurianov et al\cite{Gurianov2019BigDataFieldDevelopment} estimate that ‘big data technologies, predictive analysis and machine learning’ could reduce the costs of upstream activities (exploration and production) by 30\% by 2030. Once again, the mechanisms at work in bringing about such consequences are not clearly explained. Overall, digitalization is presented in the literature reviewed as a way of remaining competitive in the face of increasingly complex operations

However, according to Saputelli et al\cite{saputelli2024success}, few digital technologies have reached the necessary maturity in the O\&G industry over the past 20 years. The authors suggest that cloud computing, regulatory control, and wellhead sensors are the only mature technologies in the industry in 2024, while deep learning, UAVs and drones, Advanced Process Control (APC), downhole sensors and Interval Control Valves (ICV) are currently being scaled up\cite{saputelli2024success}.

The literature review reveals a contradiction. There is a significant number of scientific publications and industry reports extolling the benefits of digitalization in the O\&G sector and the various gains that can be achieved in different forms: cost reductions, increased productivity, etc. However, publications analyzing the integration of these solutions into the sector’s industrial practices indicate a low level of maturity. The IOA analysis reveals significant and growing economic flows from ICT to O\&G, far greater than those to R\&D, but the literature review suggests that very few digital products and services are reaching their full potential in this industry.

\subsection{High-level carbon footprint assessment of case studies}

As noted in the classification proposed in the Table  \ref{tab:digital_oil_gas}, digital technologies are being rolled out in environments that are not yet fully mature in order to achieve outcomes that are more or less defined. In this part of the analysis, we attempt to show how these outcomes could crystallize into potential added emissions. We have selected two cases where the expected outcomes and the mechanisms for achieving them are clear to facilitate an exploratory assessment. These two cases are situated at two different points in the value chain: one upstream and one downstream.

\subsubsection{Example 1: Microsoft Azure for XTO Energy (ExxonMobil)}

In February 2019, Microsoft announced a partnership with XTO Energy, a subsidiary of ExxonMobil, specializing in shale oil in the Permian Basin, a 1.6 million acres site with a 9.5 billion oil-equivalent barrel resource base\cite{microsoft2019exxonmobil}. The technology company supplies sensors on wellheads (temperatures, pressures, and flow rates) and Edge IoT gateways to upload data to cloud servers for real-time data analysis. Microsoft Dynamics 365 is also provided to aggregate data and provide dashboards to manage the maintenance and efficiency of hundreds of wells. Microsoft claims that XTO Energy's use of its solutions has the potential to increase daily barrel production by 50,000 by 2025\cite{microsoft2019exxonmobil}. XTO Energy had planned to extract 1 million barrels per day in the Permian Basin in 2024\cite{XTO_Permian_Basin_Operations}. Microsoft's solutions should therefore theoretically represent a 5\% increase in production. Using the proposed classification, the characteristics of this example are provided in the Table \ref{tab:microsoft-xto}.

\begin{table}[h]
\centering
\footnotesize
\renewcommand{\arraystretch}{1.25}
\setlength{\tabcolsep}{6pt}

\begin{tabularx}{\columnwidth}{lX}
\toprule
\textbf{Category} & \textbf{Description} \\
\midrule

\textbf{Activity} & Upstream; Operations \\

\textbf{Sub-activity} & Production \\

\textbf{Function} & Drilling optimization; predictive maintenance \\

\textbf{High-level service/hardware} & Sensors; networks; cloud computing \\

\textbf{Intended effect} & Increased production (5\%); cost reduction; downtime reduction \\

\textbf{Potential added emissions per year} & 6.7 MtCO$_2$ \\

\bottomrule
\end{tabularx}

\captionsetup{skip=8pt}
\caption{Microsoft and XTO Energy example classification}
\label{tab:microsoft-xto}
\end{table}

Based on a linear extrapolation from 2019 to 2025 with a 15\% average downtime, accounting for 50,000 additional barrels per day reached only in the final year, we can assume that Microsoft's digital technologies could enable the production of an additional 54,293,750 barrels. Given that a 42-gallon barrel of conventional oil has an average footprint of 431.87 kgCO2 according to the EPA\cite{EPA-GHG-Equivalencies}, the potential additional emissions from 2019 to 2025 would amount to 23,447,842 tCO2, and 6,699,383 tCO2 only for 2025. This estimate is conservative; emissions are likely to be higher because shale oil production requires more energy for extraction than conventional oil\cite{Brandt2010_CO2_OilShale}.

In 2024, Microsoft's carbon footprint was 15,543,000 (market-based) or 25,095,511 tCO2e (location-based)\cite{Microsoft2025factsheet}. The yearly additional emissions of this single project for XTO Energy (6,7 MtCO2) would then represent 43.1\% (market-based) or 26.7\% (location-based) of Microsoft's total 2024 emissions. These types of projects are not generally included in the company's Scope 3.

\subsubsection{Example 2: Downstream solutions for Valero refineries}

Gulati et al present the integration of cloud and deep learning solutions applied to downstream refinery operations\cite{gulati2020refinery}. In the case of Astron Energy in South Africa, these solutions streamline crude analysis and reduce the time required to create assays from laboratory samples by 50\%. Gulati et al estimate that this digital deployment has saved between \$20 million and \$150 million per year\cite{gulati2020refinery, AVEVA_AstronEnergy_Case}. The authors also present other use cases for their solutions to improve processes and collaboration between employees at Shell, Valero, Singapore Refining Co (SRC), and Abu Dhabi National Oil Company (ADNOC) plants. These additional savings are estimated at between \$24 million and \$135 million per year\cite{gulati2020refinery}. The authors even estimate that implementing all of these functions in a refinery could save between \$50 million and \$300 million per year. Using the proposed classification, the characteristics of this example are provided in the Table \ref{tab:aveva}.

\begin{table}[h]
\centering
\footnotesize
\renewcommand{\arraystretch}{1.25}
\setlength{\tabcolsep}{6pt}

\begin{tabularx}{\columnwidth}{lX}
\toprule
\textbf{Category} & \textbf{Description} \\
\midrule

\textbf{Activity} & Downstream; Operations \\

\textbf{Sub-activity} & Refining \\

\textbf{Function} & Production optimization; supply chain monitoring; data management; advanced process control optimization \\

\textbf{High-level service/hardware} & Sensors; networks; cloud computing \\

\textbf{Intended effect} & Cost reduction; downtime reduction \\

\textbf{Potential added emissions per year} & 0.058–3.5 MtCO$_2$ \\

\bottomrule
\end{tabularx}

\captionsetup{skip=8pt}
\caption{AVEVA example classification}
\label{tab:aveva}
\end{table}

If we take these claims at face value, we can determine the orders of magnitude of carbon emissions associated with such savings for the downstream industry. Using the method detailed in ITU-T Supplement 54\cite{ITU_LSup54_2022}, we can determine a broad monetary emissions factor from the financial results and environmental statements of companies in the sector. As an example, we take Valero, the world's largest independent refiner, as a reference point. We chose this company because it has virtually no activities other than refining, therefore its results are not mixed with other parts of the industry. Based on its after-tax profit (\$2.76 billion in 2024\cite{Valero_ARS_Combo_Book_2025}) and the carbon emissions from the company's activities (32.2 MtCO2e\cite{Valero_CY24_Assurance_Statement_2025}) in the same year, it is possible to obtain a simplified emissions factor, in this case: 11,667 kgCO2e/k€. This factor only takes into account the company's Scope 1 and 2 emissions and therefore only corresponds to refining operations. If we apply this factor to the potential savings associated with integrating Aveva's cloud solutions at Valero (\$5 to \$50 million)\cite{gulati2020refinery}, then the additional emissions would be 58 to 583 ktCO2e. Applied to all potential savings per year for one plant identified by Gulati et al\cite{gulati2020refinery}, then the additional emissions would be between 583 and 3,500 ktCO2e per year.

\section{Discussion}

 Input-Output Analysis makes it possible to define the broad orders of magnitude determining the economic relationship between the ICT sector and the O\&G sector. The quantity of inputs flowing from the ICT sector to the O\&G sector is relatively low, averaging 2\% over the period studied. This figure means that of the economic flows generated by the ICT sector, 2\% go to the O\&G sector and 98\% to other sectors, including the ICT sector itself. This does not represent a significant flow in relative terms and shows that only a small proportion of what the digital sector sells goes to the O\&G sector, mainly via telecommunications activities and the sale of IT services and consultancy. The large proportion accounted for by telecommunications is unusual and may be due to the inclusion of consultancy and IT services within this category. Overall, economic flows to the O\&G sector have a much higher carbon footprint than flows to other sectors, so 2\% of economic flows may represent significantly higher emissions relative to the scale of flows within the ICT sector. An environmentally-extended IOA would be necessary to clarify this point.
 
 In absolute value, the inputs have doubled as shown in the figure \ref{fig:inputs_totals_00_22} (20 billion dollars in 2000 and 48 billion in 2022). Overall, this value remains relatively low but only corresponds to the share of ICT that is directed specifically towards oil and natural gas production and refining, and that can be captured through IOA. The sharp increase in ICT inputs to O\&G between 2009 and 2012 can be partly explained by the development of shale oil extraction in the United States and a less favorable market, as explained by \cite{al2023review}. However, contextual elements are still lacking to reconstruct a comprehensive history of these changes in economic flows.

 Thus, a comparison focused on flows from ICT to other energy production allows us to understand the co-evolution of ICT with fossil fuels and less carbon-intensive energies such as renewable and nuclear power. As indicated in \ref{fig:inputs_totals_00_22}, there is a significant discrepancy in O\&G sector’s input from ICT, compared to that of the R\&N energy sector. This should be considered with the added information that cannot be captured by IOA: in recent years, investment in renewable energy has far surpassed that of fossil fuels (\$2.2 trillion compared to \$1.1 trillion in 2025 \cite{IEA2025WorldEnergyInvestment}). The result is even more striking while considering that a large section of the renewable energy sector are inherently ICT dependent. Even then, the O\&G industry's input from ICT sector is nearly four times higher than renewable and nuclear energy sector.

Our analysis indicates an economic dependency of the O\&G sector on ICT infrastructure. The consistent ICT spending, far surpassing that of the R\&N sector, wouldn’t have happened unless O\&G industry benefited to some extend from the digital tools offered to them. The literature review showed that the digital transformation of O\&G industry started for more than 40 years with the rise of ‘digital oilfields’, cloud computing and high-power computing (HPC) for large-scale simulation environments\cite{al2023review, saputelli2024success}, yet only few digital technologies reached maturity in this sector. However, these tools came with claims to increase discovery, recovery, and production, to improve HSE and security, or to reduce cost and downtime. Then, it's probably no surprise that the fall in oil price in 2014 led to more ICT spending rather than less, in an attempt to increase profitability by reducing costs or increasing production.

It is important to understand that the IOA does not demonstrate a causal link between the purchase of digital products and services and increased fossil fuel production. It shows that the O\&G sector is purchasing increasing amounts of ICT, but we cannot determine precisely for what purpose or what the actual effects of such flows are. However, this initial analysis, combined with the literature review and case studies, highlights a clear relationship and the risk posed by potential additional emissions resulting from digitalisation. This risk underscores the urgency of establishing whether or not a causal link exists and determining the scale of the additional emissions. 

One should highlight that O\&G industry doesn't appear to be a major source of overall revenue for the US/China-based ICT companies. In light of environmental crises and national and international carbon reduction policies, ICT companies could have divested from the O\&G sector without any severe consequences for the sector's economy. In fact, with scant data and understanding of these added emissions, the ICT industry appears to remain an appealing sector for ESG (Environmental, Social and Governance) investment\cite{bloomberg2026futureESG}. Yet major tech companies show no willingness to withdraw their resources from enabling more emissions. In fact, the boom in generative AI seems to be reigniting talk of the potential benefits to be gained from this new technology. Similarly, the increasingly prominent presence of O\&G industry players at major AI events, such as the AI Summit in Paris\cite{Khowaiter2025AdoptAI}, suggests that the economic relationship between ICT and O\&G is set to intensify.

By looking at talks at CERAweek \cite{Dang2025AI}, one of the largest gatherings of O\&G executives, one can easily understand the excitement in AI-assisted enhanced fossil fuel extraction. The recent Wood Mackenzie report \cite{WoodMackenzie_AI_Oilfield_Potential_2025}, boldly titled 'Every Last Drop: Using AI-Powered Analysis to Find Oil-Field Upside Potential,' is a concrete depiction of claims that could lead to ‘enabled emissions’. The additional emissions of accelerated O\&G exploration could be one of the most long-lasting climate impacts from AI. 

Wood Mackenzie's, estimated extraction of between 470 and 1,000 billion additional barrels from existing oil fields by 2050 would translate into an average of between 18 and 38 billion additional barrels per year. Assuming optimistically that the carbon footprint of these barrels is similar to that of a barrel of conventional oil in 2024, this would imply additional emissions of between 7.8 and 16.6 GtCO2e per year. These estimates are unrealistic, or even absurd, but reflect the renewed technological promise offered by current developments in AI. However, it is likely that oil and gas companies will increase their investment in AI in an attempt to boost productivity and reduce costs. The actual effects of these investments are impossible to assess at this stage.

\subsection{Exploring causality of ICT in O\&G sector: GPUs and simulation}

As mentioned earlier, Input-Output analysis does not allow us to establish a causal link between economic flows. The examples of XTO and Valero provide a snapshot of the links between digitalisation and increased fossil fuel production, but do not allow us to establish a general causal link between these two sectors. Establishing this causal link requires a much more detailed historical and economic analysis than we are proposing here. We wish, however, to explore how the multiple causal links between digital hardware and services and the O\&G sector might have emerged or have emerged. By way of example, we are exploring below the historical links between GPUs – through their leading manufacturer, Nvidia – and the development of seismic and reservoir simulations in the O\&G sector.

Until late 2000s, GPUs were primarily used for the niche application of graphics processing in video games with two  leading manufactures AMD (e.g., ATI-Radeon) and Nvidia (e.g., GeForce)\cite{Singer2020GPUHistory}. In 2006, Nvidia released CUDA, a so-called GPGPU (General-Purpose GPU), that can be easily programmed for other purposes. O\&G companies leveraged GPU capabilities very early on \cite{Morgan2017OilGasGPU}, and Nvidia eagerly promoted the benefits to these clients \cite{Nguyen2008CUDA}. It should be noted that “one of the first applications for CUDA was in Oil and Gas exploration, processing the mountains of data from geological surveys” \cite{Sayer2023Nvidia}.  

GPUs provided excellent efficiency gains in Seismic simulation and Reservoir analysis, very important to understand the best way of extracting hydrocarbon from existing O\&G fields and identify new ones. Long before the AI gold rush, GPU-based O\&G exploration helped creating ‘digital oilfields’ \cite{Achmad2017DOF} with HPC clusters claiming to reach the first ever billion-cell engineering calculation \cite{Natoli2017Echelon}. Software developed by Stone Ridge Technology and other companies claimed to use Nvidia GPUs to achieve several breakthroughs with far larger numbers of cells, where the O\&G industry was most often the primary commercial customer \cite{Vidyasagar2019GPUReservoir}. 

One hypothesis is that the early adoption of GPUs in such a lucrative commercial space itself provided the financial boost to develop components that could be used outside the niche gaming market. This can be seen in Nvidia’s growing dominance over AMD/ATI (both of which had roughly equal GPU market share until mid-2000 \cite{Douglas2026GPUMarket}) and its increasing use in non-gaming sector. In fact, by early 2020, Nvidia’s revenue between gaming and non-gaming sectors became roughly equal, followed by an exponential rise in its datacenter (a generic non-gaming sector) revenue \cite{Statista2025NvidiaRevenue}. Although more in-depth analysis is necessary to determine a concrete correlation, it is highly likely that these long-standing links between the two industries will be significantly amplified by the AI boom, with real implications for productivity in the O\&G sector.

\section{Limits}

We have only parsed the EXIOBASE3 Input-Output tables for transactions from regional sector to regional sector, in order to present the main macroeconomic tendencies in the relationship between ICT and O\&G at a global level. Our main focus was getting available data to feed historical trends. This is part of an exploratory work that could benefit from parsing product to product Input-Output tables, to show and analyze the material flows between sectors. In the same line of thought, EXIOBASE3 not only include monetary transactions, but also environmental extensions for the Input-Output model. Analyzing the associated emissions from sector to sector would give a better understanding of the environmental impact of the economic relationship between ICT and O\&G. 

EXIOBASE3 employs the Statistical Classification of Economic Activities in the European Community, Revision 1.1\cite{NACERevision1}, as a methodological framework for the systematic categorization of sectors. This standard was itself based on the United Nations International Standard Industrial Classification of All Economic Activities, Revision 3 (ISIC Rev. 3). The construction of a consistent database over time necessitates the maintenance of uniform categorization of economic sectors. However, this also implies that the available data may appear deficient.


Consequently, only broad sectoral categories are available, which complicates the determination of the composition of inputs from ICT to the O\&G sectors. To obtain a more comprehensive understanding of the material flows between sectors, it is necessary to complete our assessment of the economic trends from 2000 to 2022 using the data from the product-to-product transactional tables. However, this approach would not provide a clearer understanding of the services used as inputs in the O\&G sectors.


The utilization of EXIOBASE3 transaction tables facilitates the acquisition of readily available empirical data at the macroeconomic level. In order to examine the processes employed by the O\&G sectors that have their origins in the ICT sectors, it is necessary to have access to a comprehensive historical and economic record of this relationship. However, this is not sufficient on its own. A more qualitative approach is necessary for the practical implementation of inputs from the ICT sectors across the value chain in the fossil industry.


We must underline that the available data on the O\&G sector includes oil and gas extraction, oil refining and gas manufacturing, and the production of electricity from those two primary sources of energy. Concerning the R\&N energy sectors, these mostly include the production of electricity from these sources, the mining of uranium and thorium ores, and the processing of nuclear fuel are documented as sectors outside of the production of electricity.

Another limit is the granularity of available data. The sectors documented in EXIOBASE3 Input-Output tables are rather general, dependent on national accounting institutions and their labeling of economic activities. A more grounded economic and historical analysis of the relationship between ICT and O\&G (one we are aiming to contribute to) will need more detailed data and more research with sources from institutions, companies, etc, along with the macroeconomic indicators from Input-Output tables.

\section{Conclusion}
This article provides an initial overview of the economic co-evolution between the ICT and O\&G sectors, and the potential environmental consequences thereof. The first part uses Input-Output Analysis to determine the economic flows between the two sectors from 2000 to 2022. This revealed that there was absolute growth in economic flows from ICT to O\&G, which was much greater than economic flows from ICT to other energy industries, such as renewable and nuclear energy. In the second part, we categorized ICT activities based on their application within the O\&G industry's value chain and the intended outcomes. Two case studies were used to illustrate this classification, and also to define the orders of magnitude of added emissions from the digitalization of O\&G activities at different levels of the value chain. Finally, we examined the long-term effects of co-evolution between the ICT and O\&G sectors by detailing the development of GPUs in hydrocarbon reservoir simulation. We contextualized this relationship with regard to the current development of AI and its associated efficiency promises in the O\&G sector. This article highlights the urgent need for a better understanding of, and estimation of, the additional emissions caused by the digitalisation of the O\&G sector, in order to dispel the current uncertainty surrounding this issue.
\section*{Acknowledgment}

We would like to thank Francis Charpentier for his guidance on the use of Input-Output Analysis in the digital sector.
\bibliographystyle{ieeetr}
\bibliography{bibliography.bib}{}

\end{document}